\documentclass[conference]{IEEEtran}
\IEEEoverridecommandlockouts

\usepackage{cite}
\usepackage{amsmath,amssymb,amsfonts}
\usepackage{graphicx}
\usepackage{textcomp}
\usepackage{xcolor}
\usepackage{booktabs}
\usepackage{algorithm}
\usepackage{algpseudocode}
\usepackage{subfig}
\usepackage{makecell}
\usepackage{multirow}
\usepackage{diagbox}
\usepackage{pifont}
\usepackage{enumitem}
\usepackage{xurl}
\usepackage{tabularx}

\begin{document}

\title{Ding-Dong Ditch: Peeking Into Spot Instance Availability}

\author{\IEEEauthorblockN{Kyumin Kim}
\IEEEauthorblockA{\textit{Dept. of Data Science} \\
\textit{Hanyang University}\\
Seoul, Republic of Korea \\
okkimok123@hanyang.ac.kr}
\and
\IEEEauthorblockN{Moohyun Song}
\IEEEauthorblockA{\textit{Dept. of Artificial Intelligence} \\
\textit{Hanyang University}\\
Seoul, Republic of Korea \\
moohyunsong@hanyang.ac.kr}
\and
\IEEEauthorblockN{Taeyoon Kim}
\IEEEauthorblockA{\textit{Dept. of Data Science} \\
\textit{Hanyang University}\\
Seoul, Republic of Korea \\
tykim7@hanyang.ac.kr}
\and
\IEEEauthorblockN{Kyungyong Lee}
\IEEEauthorblockA{\textit{Dept. of Data Science} \\
\textit{Hanyang University}\\
Seoul, Republic of Korea \\
kyungyong@hanyang.ac.kr}
}

\maketitle


\begin{abstract}
Spot instances offer significant cost savings of up to 90\% over on-demand prices, making them an attractive resource for large-scale computing workloads. However, understanding their availability dynamics is essential for building systems that tolerate interruptions, and observing this availability directly requires keeping instances running, which incurs costs that scale with the number of monitored instance types and their per-instance price. We propose Ding-Dong Ditch (DDD), a cost-efficient method that collects spot instance availability signals by leveraging the cloud provider's provisioning lifecycle. Since the outcome of a spot request is determined before the instance enters the running state, DDD submits requests and cancels them upon provisioning acceptance, collecting binary availability signals at near-zero instance cost. Submitting multiple concurrent requests per measurement point further yields a quantitative estimate of available capacity. We validate DDD through simultaneous collection of probing signals and actual running instance traces across 68 instance types and 15 regions on both AWS and Azure, totaling 336,033 spot requests. Analysis of 2,635 real-world interruption events reveals that co-interruptions within the same instance type and availability zone occur within three minutes in over 92\% of cases, motivating a binary availability formulation. Based on this formulation, we derive three complementary features from DDD signals and demonstrate that their combination achieves an F1-macro score of up to 0.90 for current availability modeling and maintains 0.85 at a 60-minute prediction horizon. A trace-driven simulation using TPC-DS workloads further demonstrates the potential of DDD-based prediction to reduce lost computation compared to an unguided baseline.
\end{abstract}



\section{Introduction}

The growing demand for large-scale computing workloads, including machine learning, distributed data processing, and scientific simulations, has established cloud computing as a major computing platform~\cite{cloud-usage-pattern, from-cloud-to-sky, skypilot}. While the cloud can elastically provision large amounts of computing resources, the cost of such resources increases proportionally with workload scale, posing a significant challenge for cloud users~\cite{cant-be-late, parcae-nsdi}. Accordingly, cost-efficient resource utilization has been extensively studied as a key challenge in cloud-based system design~\cite{cant-be-late, interrupt-visible-www, autobot-bot-using-spot}.

From a cost optimization perspective, spot instances are an attractive option. Cloud providers offer surplus computing capacity in their data centers as spot instances at discounts of up to 90\% compared to on-demand prices, enabling users to significantly reduce their computing costs~\cite{cant-be-late, skyserve}. However, spot instances may be reclaimed when the provider's available capacity becomes insufficient or higher-priority demand arises~\cite{cant-be-late, deconstructing-spot-instance, spot-price-policy-change-2017-irwin}. Despite these constraints, the significant cost benefits have motivated a growing body of research to utilize spot instances across diverse workloads, including deep learning training~\cite{parcae-nsdi, deepspotcloud, bamboo-nsdi}, large language model serving~\cite{spotserve, skyserve}, and batch data processing~\cite{autobot-bot-using-spot, cant-be-late}.

A critical prerequisite for such systems is an accurate understanding of spot instance availability. Training jobs that span multiple nodes are sensitive to availability fluctuations, as an interruption of even a single node can force a full restart or rollback to the last checkpoint, wasting significant computation~\cite{parcae-nsdi, bamboo-nsdi}. Batch analytics pipelines must similarly anticipate capacity changes to avoid assigning long-running tasks to instances that are about to be reclaimed~\cite{cant-be-late, autobot-bot-using-spot}. The ability to predict upcoming interruptions even a few minutes in advance enables proactive strategies such as workload migration, checkpoint triggering, or query deferral, which can substantially reduce recovery overhead.

The availability of spot instances varies across instance types, regions, and over time~\cite{cant-be-late, skynomad, interrupt-visible-www}. To effectively utilize spot instances, it is therefore necessary to understand these availability patterns and establish strategies for instance type selection, region diversification, and interruption response. Cloud providers offer partial assistance through aggregate statistics such as historical interruption frequency ranges and the Spot Placement Score (SPS), an integer score reflecting instantaneous availability~\cite{spotlake-iiswc, interrupt-visible-www}. However, these datasets reflect coarse-grained trends and do not indicate whether a specific number of instances can be provisioned at a given moment. Furthermore, they lack the temporal resolution necessary to capture short-lived fluctuations that precede imminent interruptions~\cite{interrupt-visible-www}.

Given these limitations, researchers have resorted to observing spot availability directly by running actual instances and recording interruption events~\cite{spot-instance-interrupt-check-cloud-2018, interrupt-visible-www}. This approach yields high-fidelity signals under real operational conditions, but its cost scales proportionally with both the number of monitored instance types and their per-instance price. To mitigate this, Wu et al.~\cite{cant-be-late} proposed a periodic probing method that briefly launches spot instances at 10-minute intervals, reporting roughly 100$\times$ cost reduction over continuous operation. However, because probing instances are still launched and billed per invocation, the associated cost scales with the probing frequency and the number of target types, limiting its practicality for large-scale or fine-grained studies. In addition, neither continuous monitoring nor periodic probing provides a quantitative characterization of how closely the collected signals correspond to actual runtime availability, leaving the fidelity of these approaches unverified.

To address this challenge, we propose Ding-Dong Ditch (DDD), a method for monitoring the availability of spot instances across diverse instance types and regions without keeping instances running. DDD leverages a structural property of the cloud provider's provisioning lifecycle, where the outcome of a spot request is determined before the instance enters the running state and incurs compute charges. By submitting spot requests and canceling them upon acceptance of provisioning, DDD collects binary availability signals at near-zero instance cost. Submitting multiple concurrent requests at each measurement point yields a quantitative estimate of available capacity, enabling continuous observation across a large number of instance types and regions at significantly reduced cost. Beyond data collection, we analyze real-world spot interruption patterns, define availability features tailored to the observed characteristics, and present a system that integrates collection, feature computation, and prediction into a unified pipeline. The main contributions of this paper are as follows:

\begin{itemize}[leftmargin=*]
  \item We propose DDD, a cost-efficient availability observation method that leverages the cloud provider's provisioning lifecycle. DDD achieves 2.5$\times$ lower monitoring cost than periodic probing~\cite{cant-be-late} and 249.5$\times$ lower cost than continuous monitoring, while collecting data more than 3$\times$ finer temporal resolution.
  \item We conduct an empirical analysis across diverse instance types and regions on both AWS and Azure, quantitatively characterizing the relationship between DDD signals and actual running instance behavior, including the types and rates of discrepancy.
  \item We analyze real-world spot interruption patterns to formulate a practical binary availability definition and derive three complementary features from DDD signals that capture instantaneous state, sustained degradation, and state transition duration.
  \item We present a system architecture that integrates DDD data collection, incremental $O(1)$ feature computation, and interruption prediction into a unified pipeline deployable on serverless infrastructure.
  \item We demonstrate that the proposed features achieve an F1-macro score of up to 0.90 for availability modeling and maintain above 0.85 at a 60-minute prediction horizon. A trace-driven simulation further confirms the potential to reduce lost computation relative to an unguided baseline.
\end{itemize}

\section{Background and Related Work}

\subsection{Spot Instance}
Cloud providers offer idle computing resources in their data centers as spot instances~\cite{deconstructing-spot-instance,cant-be-late,cloud-usage-pattern}. This pricing model allows providers to dynamically sell surplus capacity, thereby improving overall resource utilization. Major cloud providers, including Amazon Web Services (AWS), Microsoft Azure, and Google Cloud Platform (GCP), offer their own variants of such services. Spot instances provide significant cost savings, with discounts of up to 90\% compared to on-demand prices, making them an attractive option for cost-sensitive workloads.

The defining characteristic of spot instances is the possibility of interruption. Cloud providers may reclaim spot instances at any time in response to changes in resource demand, which is the fundamental difference from on-demand instances. Although providers issue advance notifications before interruption, the notice period is typically short, ranging from 30 seconds to two minutes depending on the provider, leaving limited time for applications to react~\cite{aws-spot-instance-interruption-notice, azure-spot-vms, gcp-spot-vms}. Pricing and eviction policies also vary across providers, adding further complexity to the reliable utilization of spot instances.

\subsection{Spot Instance Utilization}
Despite the cost advantages of spot instances, the possibility of interruption threatens the continuity of workload execution. Accordingly, various system-level studies have been conducted to utilize spot instances reliably and efficiently~\cite{bamboo-nsdi,deepvm-ccgrid,snape-azure-spot-mixture,parcae-nsdi,spotserve}.

Checkpointing-based approaches periodically save the state of tasks running on spot instances, enabling resumption from the last checkpoint when an interruption occurs. This is effective in minimizing computational loss due to interruptions. Bamboo~\cite{bamboo-nsdi} proposed a cost-efficient DNN training system that leverages redundant computation and checkpointing to maintain training progress on preemptible instances. DeepVM~\cite{deepvm-ccgrid} presented a system that integrates spot and on-demand VMs for cost-efficient deep learning clusters, employing checkpoint-restart mechanisms to maintain training progress when spot instances are reclaimed.

Diversification strategies across multiple instance types and regions have also been actively studied. Relying on a single instance type makes a workload vulnerable to availability fluctuations of that type. To mitigate interruption risk, approaches have been proposed that distribute workloads across multiple instance types and regions or mix spot and on-demand instances. Parcae~\cite{parcae-nsdi} proposed an adaptive resource reconfiguration technique that proactively adjusts parallelization configurations to handle spot interruptions during deep learning training. Snape~\cite{snape-azure-spot-mixture} presented an intelligent framework that dynamically mixes spot and on-demand VMs using constrained reinforcement learning, achieving significant cost savings while maintaining high availability. SpotServe~\cite{spotserve} addressed the challenge of serving generative large language models on preemptible instances by dynamically adapting parallelization configurations and introducing stateful inference recovery to minimize the impact of interruptions.

These studies share a common prerequisite of understanding the availability characteristics of spot instances. Effective utilization strategies depend on knowledge of which instance types remain stable in which regions and how interruption patterns evolve over time. However, cloud providers do not provide real-world availability data, and observing it independently requires running actual instances, which incurs substantial operational overhead.

\subsection{Monitoring Spot Availability}

Cloud providers have begun to expose datasets that reflect spot instance availability. AWS and Azure publish the Spot Placement Score (SPS), an integer score indicating current availability. They also provide a categorical interruption frequency range for each instance type.  SpotLake~\cite{spotlake-iiswc} proposed a system for continuously collecting and archiving spot instance datasets from multiple cloud providers, enabling public access to historical records. Kim et al.~\cite{interrupt-visible-www} analyzed these datasets and demonstrated that the SPS dataset exhibits periodic change patterns at hourly and daily granularity, showing significant correlation with actual spot instance interruption events. 
However, these provider-supplied datasets indicate relative availability trends but do not guarantee whether a specific number of instances of a given type can be reliably obtained at a particular point in time. Furthermore, they lack the temporal resolution necessary to capture short-lived availability fluctuations.

Given these limitations in publicly available data, researchers have resorted to running actual spot instances and recording interruption events directly~\cite{spot-instance-interrupt-check-cloud-2018, interrupt-visible-www}. This approach enables observation under conditions identical to real operating environments. However, the cost of this method scales proportionally with the number of monitored instance type and region combinations, as each instance must remain running throughout the observation period. For long-term monitoring across a large number of instance types, the associated cost becomes a practical barrier to research.

To reduce this cost, Wu et al.~\cite{cant-be-late} proposed a periodic probing method that briefly launches spot instances at regular intervals and infers availability from the success or failure of requests. This approach achieves monitoring at approximately one-tenth the cost of continuous operation. The authors visually compared their probing results with actual spot interruption events and concluded that probing signals sufficiently represent real preemption patterns, citing visual similarity and qualitative claims of high correlation. However, no quantitative metrics were provided to measure how strongly spot request outcomes correlate with actual runtime interruptions. The analysis also did not examine what types of discrepancies, such as false positives or false negatives, exist between the two signals. Moreover, since instances are actually launched and billed at each probing interval, the cost still grows with the number of target instance types and the probing frequency, limiting its practicality for large-scale studies.

In summary, the datasets published by cloud providers are accessible at no additional cost but lack the resolution for precise availability assessment, while empirical methods offer higher fidelity at a cost that scales with the number and price of target instances. To overcome this trade-off, we propose Ding-Dong Ditch (DDD), a method that collects availability signals while minimizing instance runtime costs by probing the cloud provider's provisioning process without keeping instances running. Beyond cost-efficient collection, we quantitatively analyze the discrepancies between DDD signals and actual running instance data and demonstrate how these signals can be leveraged for availability modeling and interruption prediction.

\section{Cost-Efficient Availability Collection}

DDD is a method for collecting spot instance availability signals that utilizes the structural properties of cloud providers' provisioning lifecycle. By submitting and canceling spot requests during the provisioning phase, DDD captures availability information with minimal operational overhead.

\begin{figure}[t]
    \centering
    \includegraphics[width=0.48\textwidth]{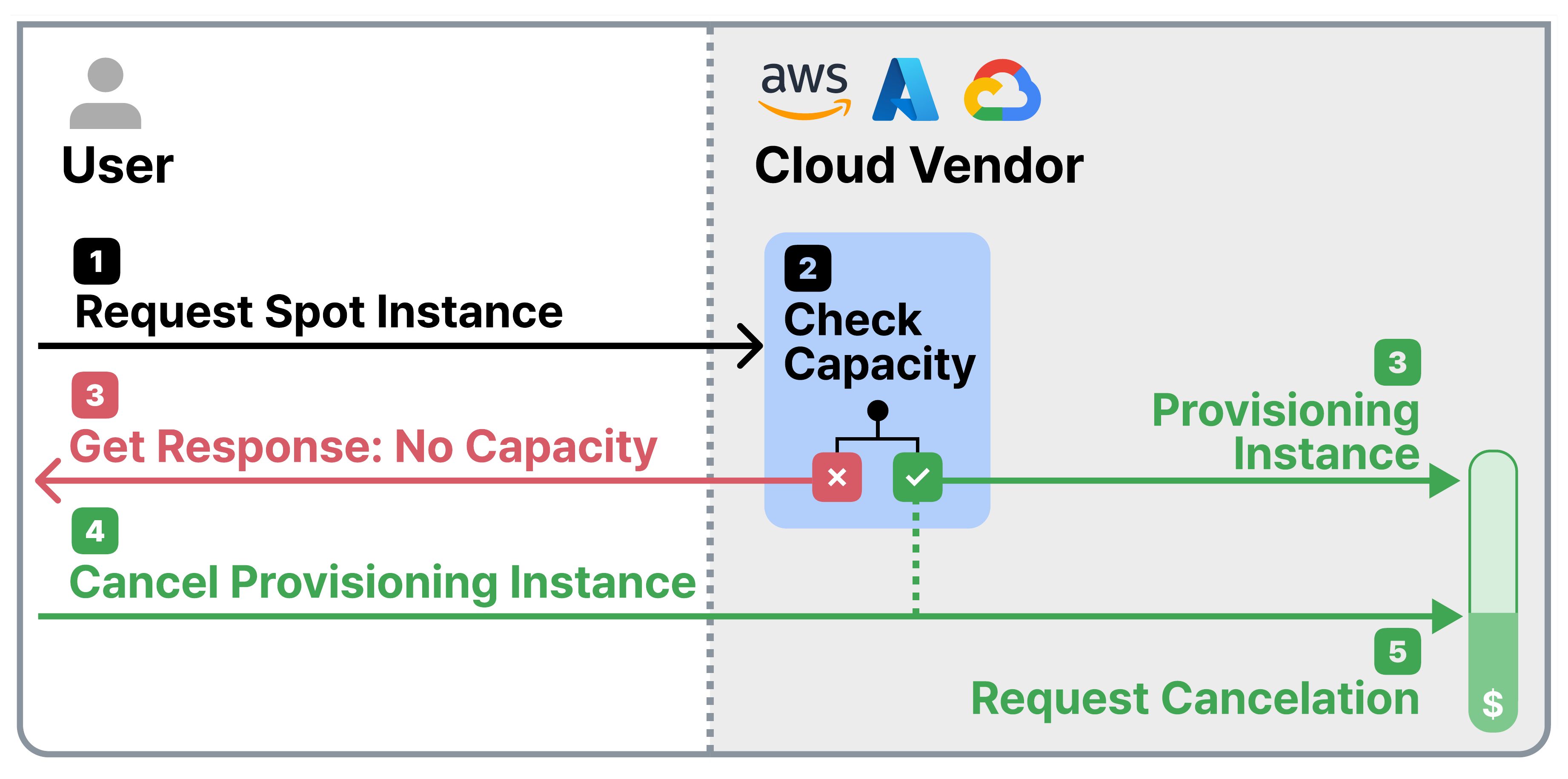}
    \caption{Spot instance request lifecycle with capacity check}
    \label{fig:spot-lifecycle}
\end{figure}

\subsection{Probing Availability via the Provisioning Lifecycle}

When a user submits a spot instance request, the cloud provider checks whether sufficient capacity is available. If capacity is insufficient, the provider rejects the request without entering the provisioning phase. Otherwise, the provider enters a provisioning phase to allocate the requested resources. During this phase, the resources are being prepared but are not yet available to the user. As provisioning nears completion, the instance transitions into the running state and becomes available to the user. Consequently, the outcome of a spot request, whether the provider can fulfill it, is determined before the instance actually begins running.

DDD operates within this pre-running window to probe spot instance availability. The name reflects this mechanism, drawing an analogy to ringing a doorbell and leaving before it is answered. DDD submits a request and observes the provider's response before the instance enters the running state. Each request yields one of two outcomes. If the provider rejects the request due to insufficient capacity, this rejection signals that the requested instance type is currently unavailable. Conversely, as illustrated in Fig.~\ref{fig:spot-lifecycle}, if the provider accepts the request and provisioning begins, DDD cancels the request upon observing the provisioning lifecycle event, before the instance enters the running state. The successful acceptance of the request signals that capacity exists for the requested instance type.

Since each request yields only a binary outcome of accepted or rejected, a single request cannot capture the degree of available capacity. To estimate how much capacity is currently available for a given instance type, DDD submits multiple concurrent requests at each measurement point. The ratio of accepted requests to the total submitted then serves as a quantitative indicator of the available capacity at that time. Because each probe completes before the instance reaches the running state, DDD minimizes the computing resource costs that account for most of the overhead in conventional spot availability monitoring. This enables cost-efficient observation across a large number of instance types and regions.

\begin{figure*}[t]
    \centering
    \includegraphics[width=0.22\textwidth]{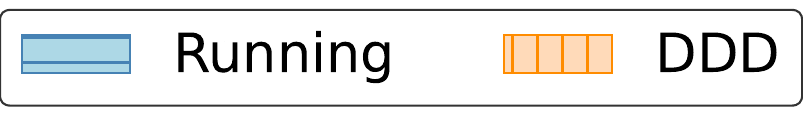}
    
    \subfloat[Synchronized recovery]{
        \includegraphics[width=0.234\textwidth]{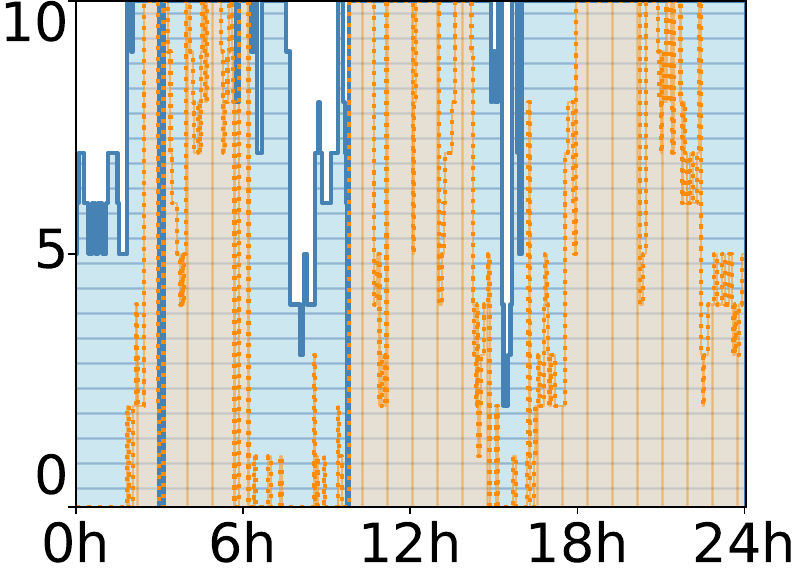}
        \label{fig:compare-case-0}
    }%
    \subfloat[DDD not recovered]{
    \includegraphics[width=0.234\textwidth]{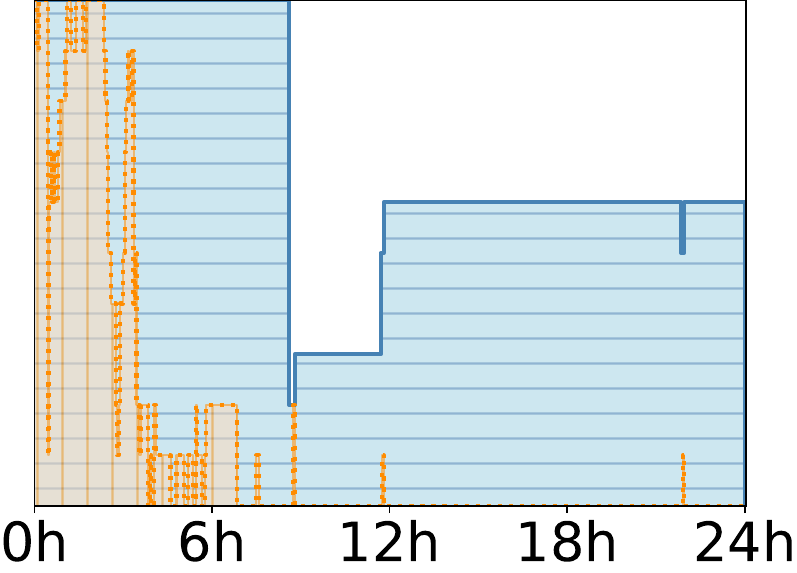}
        \label{fig:compare-case-1}
    }%
    \subfloat[Fully fulfilled]{
        \includegraphics[width=0.234\textwidth]{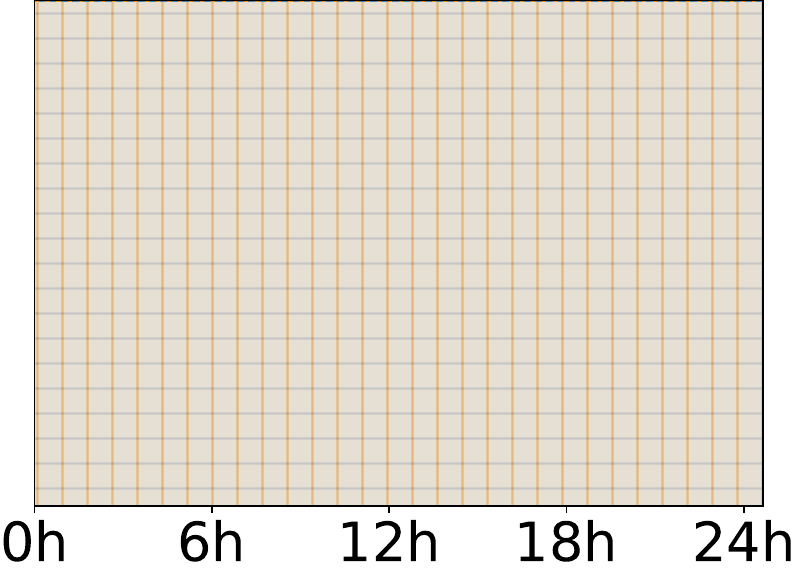}
        \label{fig:compare-case-2}
    }%
    \subfloat[DDD not fulfilled]{
        \includegraphics[width=0.234\textwidth]{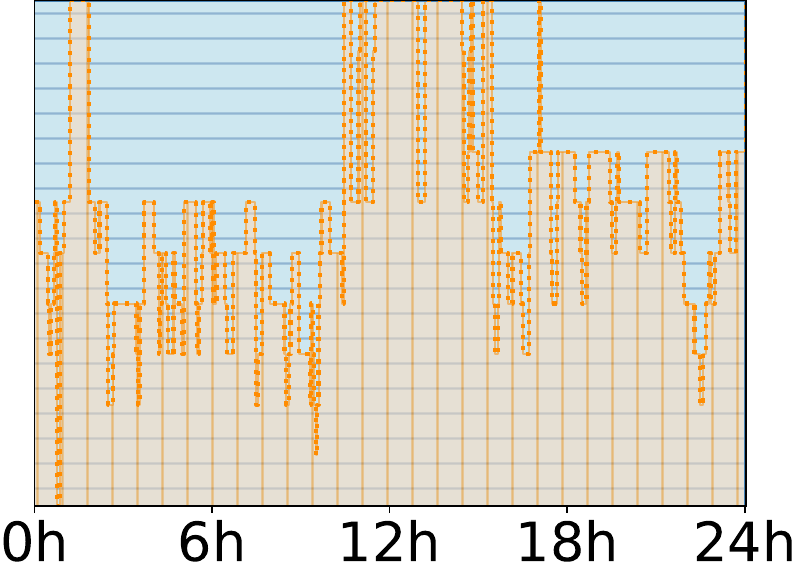}
        \label{fig:compare-case-3}
    }%
    \caption{DDD signals and actual running instance traces}
    \label{fig:ddd-checker-data-comparison}
\end{figure*}

\begin{table}
\caption{Per-time-point comparison of DDD successful request count versus actual running instance count for each instance type}
\begin{center}
\begin{tabularx}{\columnwidth}{l*{3}{>{\centering\arraybackslash}X}}
    \toprule
    \textbf{Provider} & \textbf{Actual $>$ DDD} & \textbf{Actual $=$ DDD} & \textbf{Actual $<$ DDD} \\
    \midrule
    AWS   & 4996 (22.31\%) & 17267 (77.12\%) & 126 (0.56\%) \\
    Azure & 1110 (11.03\%) & 8927 (88.68\%)  & 30 (0.30\%)  \\
    \bottomrule
\end{tabularx}
\end{center}
\label{tab:aws-azure-table}
\end{table}

\subsection{Spot Availability Characterization}\label{sec:characterization}

To validate that DDD signals reflect actual spot instance availability, we simultaneously collected DDD data and ran spot instance node pools to record real interruption events. Data collection was conducted on both AWS and Azure over multiple experiment rounds. DDD submitted requests at 3-minute intervals, the minimum period that avoids exceeding the provider's API rate limit in our experiment scenario. Each experiment simultaneously requested 10 nodes per instance type. For AWS, we conducted four 24-hour experiments in December 2024, January 2025, February 2026, and March 2026, monitoring 47 instance types across 11 regions, including major regions such as us-east-1, us-west-2, eu-west-1, and ap-northeast-1. For Azure, we conducted two 24-hour experiments in March 2026, monitoring 21 instance types across 4 regions corresponding to the major AWS regions such as US East, US West 2, Europe West and Japan East. In total, 336,033 spot requests were submitted across all experiments.

Table~\ref{tab:aws-azure-table} compares the number of successful DDD requests with the number of actual running instances at each measurement point for the same instance type. For the majority of measurement points, the two counts matched exactly, with an agreement rate of 77.12\% for AWS and a slightly higher 88.68\% for Azure. Among the non-matching cases, the actual running count exceeded the DDD success count in 22.31\% of AWS and 11.03\% of Azure measurement points, whereas the reverse case accounted for only 0.56\% and 0.30\%, respectively. This indicates that DDD rarely overestimates actual availability, suggesting its potential as a reliable indicator of actual instance availability.

Fig.~\ref{fig:ddd-checker-data-comparison} presents four representative cases from the simultaneously collected DDD and running instance traces. In each subfigure, the x-axis represents elapsed time since the start of the experiment, and the y-axis shows the number of successful DDD requests and the number of actual running instances at each measurement point.

Figs.~\ref{fig:compare-case-0} and~\ref{fig:compare-case-1} illustrate how DDD signals and running instances respond to capacity fluctuations. In both cases, the number of successful DDD requests fluctuated as available capacity changed. Fig.~\ref{fig:compare-case-0} shows a case where capacity recovered and the running instance count returned to its previous level. However, the number of successful DDD requests increased more slowly, indicating that DDD reflects capacity recovery conservatively. Fig.~\ref{fig:compare-case-1} shows a contrasting case where the running instance count recovered but the number of successful DDD requests remained lower, indicating that DDD continued to reflect reduced capacity despite the apparent recovery.

Figs.~\ref{fig:compare-case-2} and~\ref{fig:compare-case-3} show that DDD results can differ even when the number of running instances is identical. Although all requested instances remained operational in both cases, Fig.~\ref{fig:compare-case-2} exhibited a stable number of successful DDD requests matching the running instance count, whereas Fig.~\ref{fig:compare-case-3} showed continuously fluctuating values. This pattern suggests that the underlying spot availability can vary even when the same number of instances are running.

Across the collected data, we consistently observed that DDD signals reflect capacity changes that have not yet manifested as actual interruptions or recoveries. This property suggests that DDD data may serve as a basis for predicting future availability transitions.



\section{Predicting Spot Availability}

The characterization in the previous section showed that DDD signals closely track actual instance availability at most measurement points but exhibit discrepancies in a subset of cases. In this section, we analyze the patterns underlying real-world spot interruptions to inform the design of features that can bridge this gap, enabling DDD data to be used for effective availability modeling and prediction.

\begin{figure}[htbp]
    \centering
    \includegraphics[width=0.95\columnwidth]{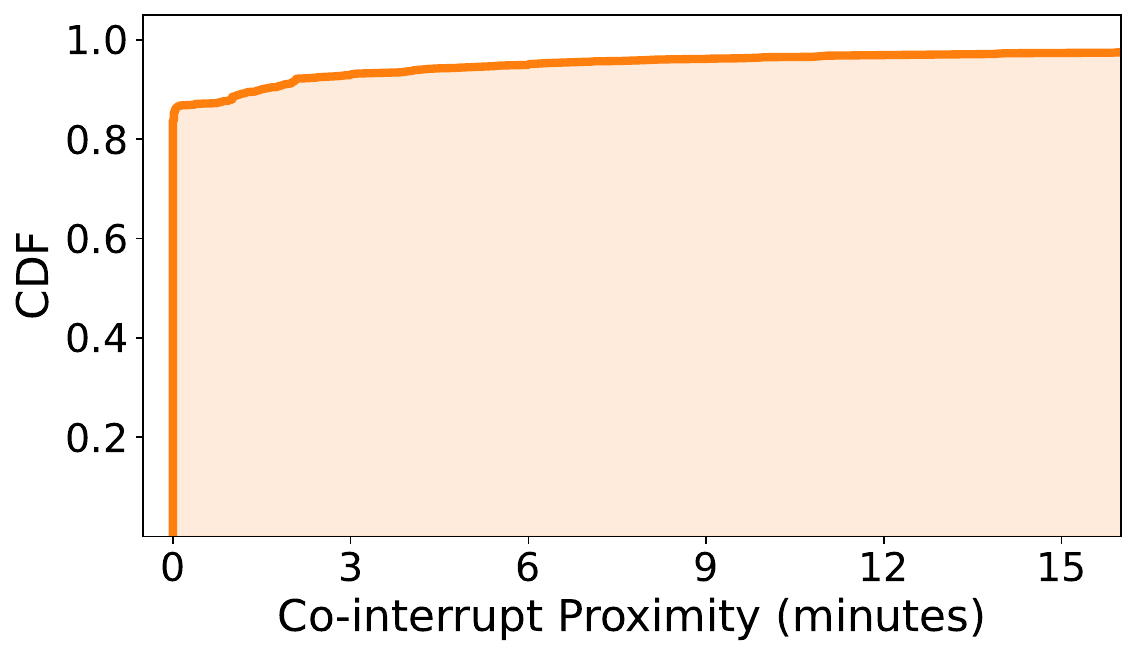}
    \caption{Cumulative distribution of co-interrupt proximity for same instance types. Most interruptions are closely followed by interruptions of other nodes of the same type.}
    \label{fig:cointerrupt}
\end{figure}

\subsection{Co-Interruption Patterns and Availability Formulation}
During the data collection, we recorded a total of 2,635 interruption events across all instance types and regions. Fig.~\ref{fig:cointerrupt} shows the cumulative distribution of co-interrupt proximity, defined as the time elapsed between an interruption of one node and the nearest interruption of another node of the same instance type in the same availability zone. Over 85\% of co-interruptions occurred within less than one minute, and 92.9\% occurred within 3 minutes. These results indicate that when any single node of a given instance type experiences an interruption, other nodes of the same type in the same availability zone are highly likely to be interrupted shortly afterward.

This pattern can be explained by the shared capacity pool structure of cloud providers. Instances of the same type within the same availability zone draw from a common resource pool. When available capacity in this pool becomes insufficient, the provider reclaims resources from the pool as a whole, causing correlated interruptions across co-located nodes. An interruption of even a single node therefore signals that the underlying capacity for that instance type has already become constrained.

Given this observation, predicting the exact number of surviving nodes at fine-grained resolution provides limited practical value. Instead, a more meaningful formulation is to determine whether the full set of requested nodes remains available or not. We therefore adopt a binary availability definition. At each measurement point, we determine whether all $N$ requested instances of a given type are fulfilled or not. This formulation captures the dominant operational pattern observed in the data and provides the foundation for the feature definitions that follow.

\subsection{Availability Features from DDD}\label{extracting-features}

Based on the binary availability formulation from the previous subsection, we define three features that capture different temporal aspects of spot availability from DDD signals. At each measurement point, DDD submits $N$ concurrent requests for a given instance type. We index each collection cycle as $t = 1, 2, \ldots$ with a fixed collection interval $\Delta t$, and denote the number of successful requests at cycle $t$ as $S_t$. Given a time window of $W$ minutes, let $w = W / \Delta t$ denote the window length in collection cycles.

\textbf{Success Rate (SR)} represents the proportion of fulfilled requests at a single collection cycle:
\begin{equation}
    SR(t) = \frac{S_t}{N}
\end{equation}
As shown in the characterization results where DDD signals matched actual running instance counts at the majority of measurement points, SR directly reflects current availability in most cases and serves as the baseline signal for availability assessment.

\textbf{Unfulfilled Ratio (UR)} quantifies the proportion of unfulfilled requests accumulated within a sliding window of length $w$. We maintain a cumulative array $P[t] = P[t{-}1] + (N - S_t)$ over unfulfilled request counts, with $P[0] = 0$. The windowed UR is then:
\begin{equation}
    UR(t, w) = \frac{P[t] - P[t{-}w]}{w \cdot N}
\end{equation}
UR accumulates failure rates within the window, so it captures sustained degradation over a recent time period. However, UR responds slowly to sudden recoveries, as past failures within the window continue to influence the value.
\textbf{Contiguous Unfulfilled Time (CUT)} measures the accumulated time during which the most recent sequence of unfulfilled states persists:
\begin{equation}
    CUT(t) = \begin{cases} 0 & \text{if } t = 1 \text{ or } S_t = N \\ CUT(t{-}1) + \Delta t & \text{otherwise} \end{cases}
\end{equation}
CUT resets to zero at the first collection cycle and whenever all requests are fulfilled, and grows by $\Delta t$ as the unfulfilled state persists. Unlike UR, CUT discards all information from before the most recent fulfilled state, capturing only the duration of the current degradation episode.

These features are complementary in the temporal range they cover. SR serves as the baseline signal that directly reflects current availability. UR and CUT supplement SR by capturing historical availability trends that a single measurement does not reflect. UR indicates whether availability has been generally stable or degraded over the past period within the window, while CUT provides responsiveness to recent state transitions. This complementarity is particularly important when $w$ is set to a larger value. As $w$ grows, UR reflects increasingly distant past trends and becomes less responsive to recent changes. CUT compensates for this by continuing to capture instantaneous state transitions, preserving information about recent availability changes that UR alone would miss.

\begin{algorithm}
\caption{Incremental Feature Update}\label{alg:feature-update}
\begin{algorithmic}[1]
\Require New measurement $S_t$, time window $W$ (minutes), collection interval $\Delta t$, cumulative array $P[\,]$
\State \textbf{Initialize:} $P[0] \gets 0,\; CUT_0 \gets 0$
\State $w \gets W \,/\, \Delta t$ \Comment{Window length in collection cycles}
\State $SR_t \gets S_t \,/\, N$
\State $P[t] \gets P[t{-}1] + (N - S_t)$
\If{$t \geq w$}
    \State $UR_t \gets (P[t] - P[t{-}w]) \,/\, (w \cdot N)$
\Else
    \State $UR_t \gets (P[t] - P[0]) \,/\, (t \cdot N)$ 
\EndIf
\If{$t = 1$ \textbf{or} $S_t = N$}
    \State $CUT_t \gets 0$
\Else
    \State $CUT_t \gets CUT_{t-1} + \Delta t$
\EndIf
\State \Return $(SR_t,\; UR_t,\; CUT_t)$
\end{algorithmic}
\end{algorithm}

All three features can be maintained incrementally with $O(1)$ computation per update using the sliding window approach described in Algorithm~\ref{alg:feature-update}. By maintaining the cumulative array $P$, which stores the running total of unfulfilled request counts up to each collection cycle, as an auxiliary data structure alongside the three features, UR can be computed from two stored values without iterating over the entire window. CUT requires only a comparison with the previous state. This computational efficiency makes the features suitable for real-time collection environments where feature updates and predictions must be produced at every cycle.

\section{DDD System Implementation}

\begin{figure*}[htbp]
    \centering
    \includegraphics[width=0.99\textwidth]{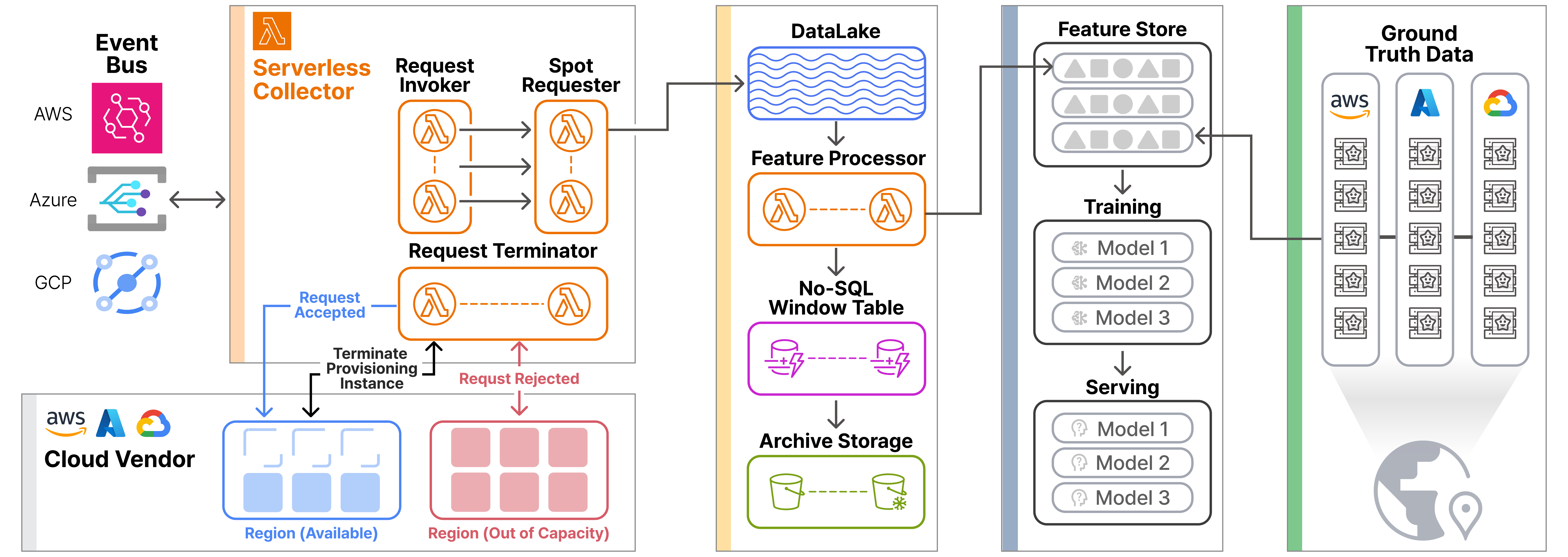}
    \caption{Architecture of the DDD system}
    \label{fig:DDD-architecture}
\end{figure*}


The DDD system consists of three modules, as illustrated in Fig.~\ref{fig:DDD-architecture}. The \textbf{DDD Collector} periodically probes spot availability across multiple instance types and regions. The \textbf{Data Pipeline} processes collection records into input features for prediction. The \textbf{Interrupt Predictor} produces real-time interruption forecasts based on the computed features.

\textbf{DDD Collector.}
The DDD Collector is implemented on a serverless architecture using AWS Lambda as the primary execution environment. This design leverages elastic scalability to perform a large number of concurrent requests within a short time window, enabling simultaneous availability measurement across multiple node pools. The Collector comprises three components. The \textbf{Request Invoker} manages the list of target instance types and periodically triggers data collection via AWS EventBridge. It is deployed on the same cloud service as the requester functions to minimize temporal discrepancies across regions. The \textbf{Parallel Spot Requester} consists of Lambda functions deployed in each target region. Each function submits a spot instance request through the cloud provider's API and records the outcome in the Data Lake. The \textbf{Request Terminator} responds to provisioning events from the cloud provider and immediately cancels the accepted request. Adopting an event-driven approach allows the Terminator to react independently of the Spot Requester, minimizing the provisioning duration and ensuring that resources are released before allocation completes.

\textbf{Data Pipeline.}
The Data Pipeline transforms records into features suitable for prediction. The \textbf{Data Lake} stores the outcome of each request as a record containing the timestamp, instance type, region, and acceptance or rejection status.
As new records accumulate, the \textbf{Feature Processor} computes features incrementally using a sliding window algorithm. The \textbf{Window Table} maintains accumulated results from recent collection cycles and serves as the computation substrate for the Feature Processor. Incremental computation ensures that each update operates with $O(1)$ time complexity by leveraging the previously computed window state, allowing feature extraction to remain fast regardless of the overall data collection scale. The resulting features are forwarded to the Interrupt Predictor for inference and simultaneously written back to the Window Table for use in subsequent cycles. The Feature Processor transfers data that falls outside the window range to the \textbf{Data Archive} for future use.

\textbf{Interrupt Predictor.}
The Interrupt Predictor takes the features of a single data point from the Feature Processor as input and predicts whether the target instance node pool will maintain its current scale over a specified future horizon. The Predictor performs inference using a pre-trained XGBoost~\cite{xgboost} model deployed as a Lambda function. As a tree-based model, XGBoost requires no specialized accelerators and runs on minimal computing resources, enabling fast and cost-efficient inference. Upon completion, the Predictor attaches the prediction result to the corresponding input record and stores it in the Window Table.

\section{Evaluation} \label{evaluation}

The preceding sections established that DDD signals closely track actual spot instance availability and that the proposed features capture complementary aspects of availability dynamics. We now evaluate these claims empirically and investigate the practical utility of DDD through the following research questions.

\begin{itemize}[leftmargin=*]
    \item \textbf{RQ1.} How does the monitoring cost of DDD compare to that of existing spot availability observation methods?
    \item \textbf{RQ2.} How accurately do features extracted from DDD data model the availability state of actual running spot instances?
    \item \textbf{RQ3.} Can DDD-derived features predict whether a spot node pool will remain fully available over a given future horizon?
    \item \textbf{RQ4.} To what extent can DDD-based availability prediction improve spot instance utilization under realistic workload conditions?
\end{itemize}

\subsection{Experiment Setup}

We evaluate DDD using the dataset described in Section~\ref{sec:characterization}, which covers diverse instance types and regions across multiple cloud providers.

To assess the generality of DDD-derived features across different learning approaches, we compare six classification models. Four models operate on a single data point per prediction such as Logistic Regression, Support Vector Machine, Random Forest, and XGBoost. The remaining two models operate on a sequence of consecutive data points: Long Short-Term Memory (LSTM)~\cite{lstm} and Transformer~\cite{transformer}. The prediction experiments vary the feature window size from 60 to 720 minutes and select the window size that yields the best performance for each model. The sequence models use an input sequence length equal to the selected window size to ensure a consistent comparison. The data is split into 75\% for training and 25\% for testing by a fixed random seed.

The F1-macro score is adopted as the evaluation metric, as it equally weights the prediction performance on both fulfilled and unfulfilled states.

\begin{figure*}
    \centering
    \begin{minipage}[t]{0.26\textwidth}
        \centering
        \includegraphics[height=10em]{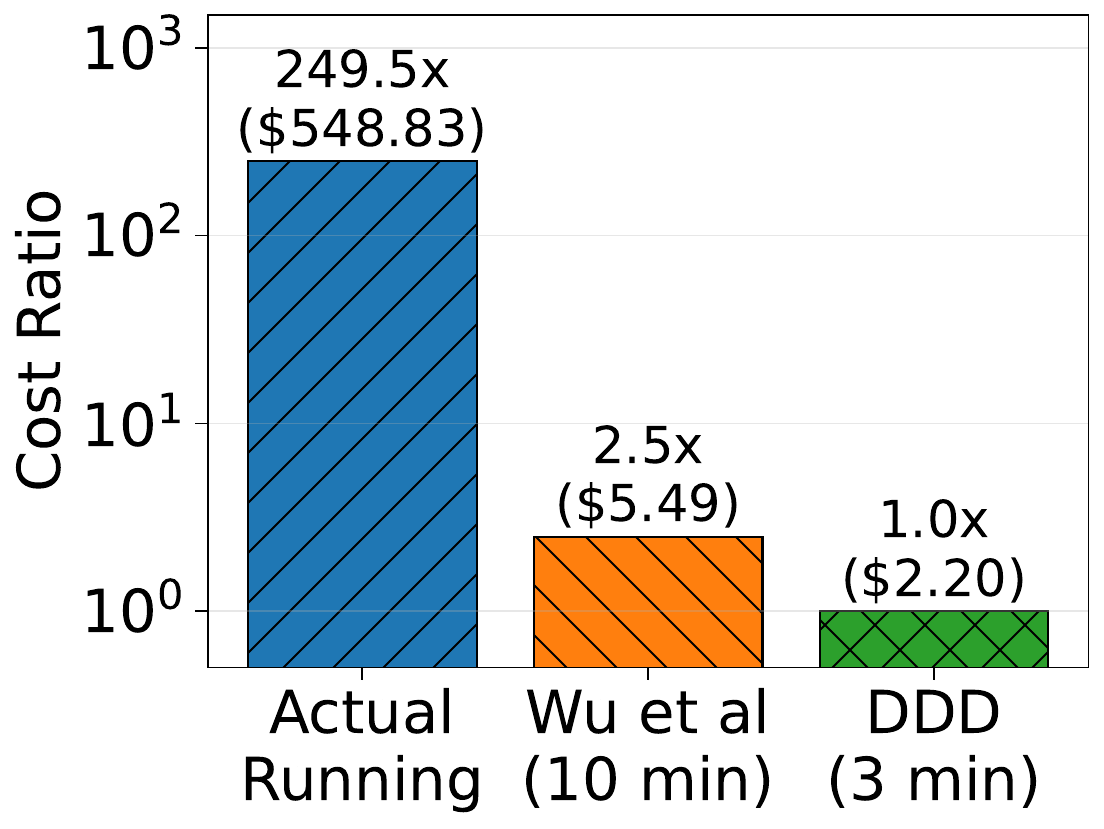}
        \vspace{0.1em}
        \caption{24-hour spot instance monitoring cost comparison}
        \label{fig:cost-comparison}
    \end{minipage}%
    \hfill
    \begin{minipage}[t]{0.71\textwidth}
        \centering
        \subfloat[Success Rate]{%
        \includegraphics[height=10em]{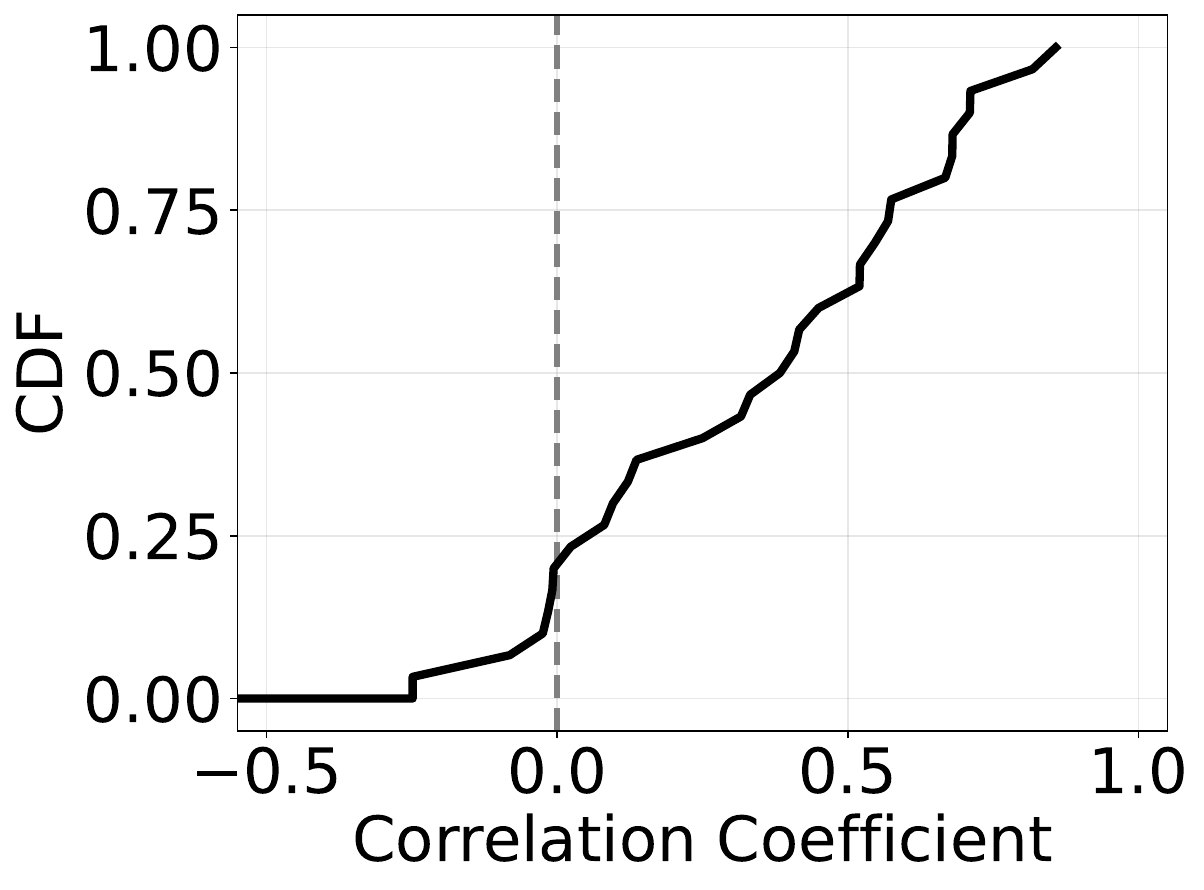}%
            \label{fig:cdf-SR}
        }%
        \subfloat[Unfulfilled Ratio]{%
            \includegraphics[height=10em]{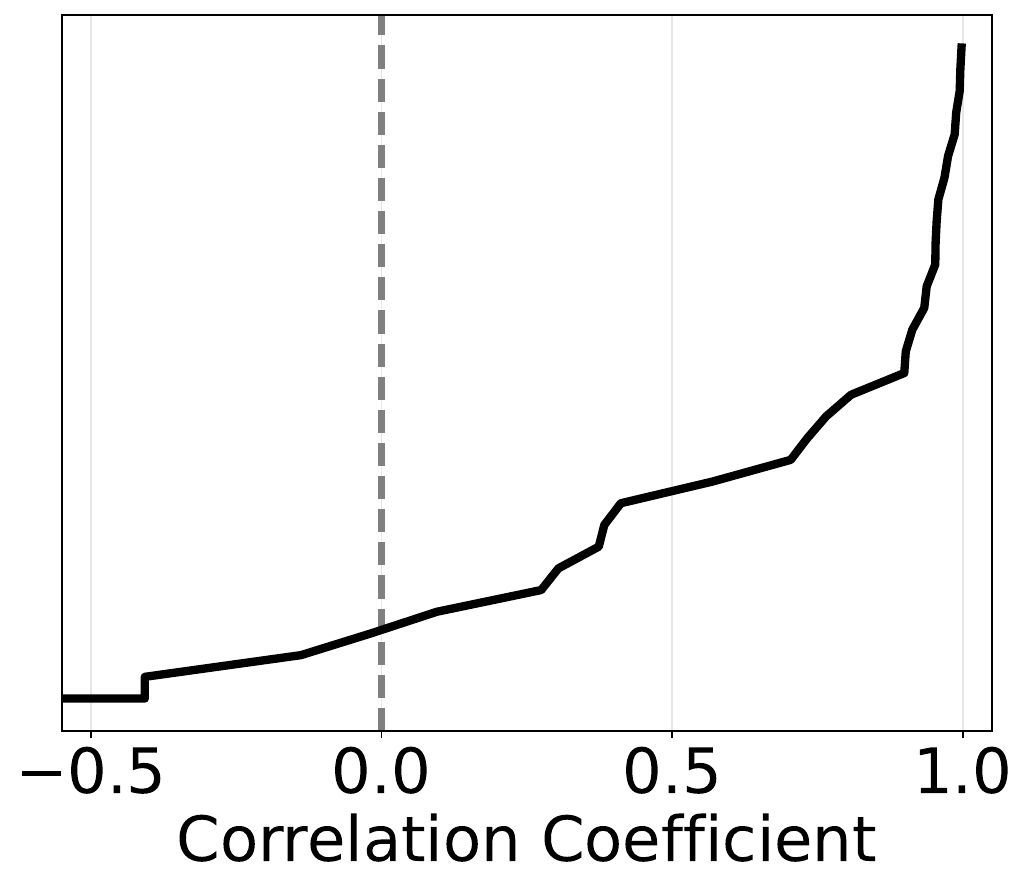}%
            \label{fig:cdf-UR}
        }%
        \subfloat[Contiguous Unfulfilled Time]{%
        \includegraphics[height=10em]{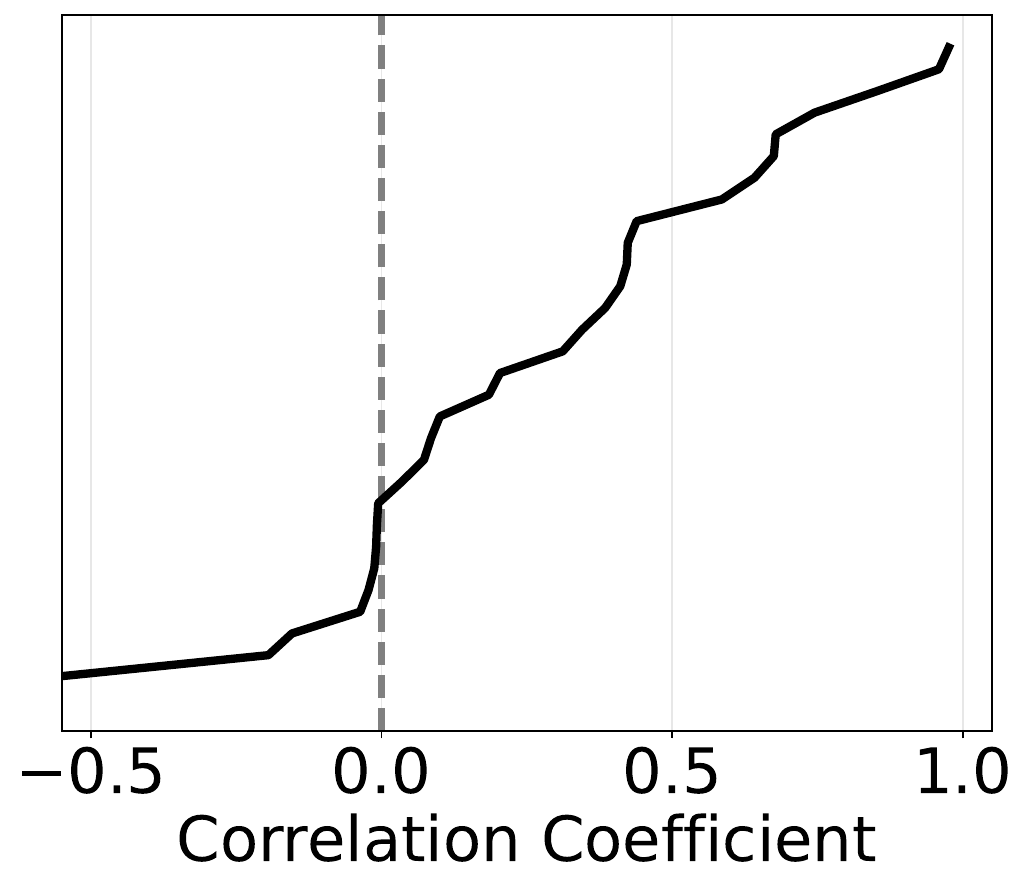}%
            \label{fig:cdf-CUT}
        }%
        \caption{CDFs of Pearson correlation coefficients between DDD-derived and actual-instance-derived features across instance types}
        \label{fig:feature-fidelity}
    \end{minipage}
\end{figure*}

\subsection{Cost Efficiency}


Wu et al.~\cite{cant-be-late} proposed a periodic probing method that launches spot instances every 10 minutes to collect availability data, reporting approximately 100$\times$ cost reduction compared to continuously running instances on AWS. However, the specific implementation that achieves this reduction is not described in the original work. We reimplemented the procedure for direct comparison but did not reproduce the reported level of cost reduction. We therefore adopt the reported 100$\times$ reduction ratio, the upper bound claimed for this approach, as the comparison baseline.

Fig.~\ref{fig:cost-comparison} compares the costs of three monitoring approaches based on the actual expenditure incurred during the four rounds of AWS data collection conducted in this study. The y-axis is shown on a logarithmic scale, with all values normalized to the cost of DDD. Continuously running spot instances throughout the collection period would have cost approximately 249.5$\times$ the DDD cost. Applying the 100$\times$ reduction reported by Wu et al. yields an estimated cost that is still 2.5$\times$ higher than DDD. This cost advantage results from the provisioning-event-based collection design of DDD. Each collection cycle requests a target instance, records the provisioning outcome, and releases the resource within seconds, minimizing charges for the target compute resources. Consequently, whereas traditional approaches incur most of their cost from the compute resources being monitored, the majority of DDD's operating cost arises from serverless collector invocations and log storage, resulting in lower overhead.

Notably, DDD achieves this cost reduction while collecting availability data at approximately 3$\times$ higher temporal resolution than Wu et al., probing every 3 minutes compared to their 10-minute interval.

\subsection{Feature Fidelity}

To verify that features extracted from DDD data faithfully reflect the patterns observed in actual running instance data, we compute the Pearson correlation coefficient for each feature between the two data sources. For each instance type, we apply the feature extraction method described to both the actual availability trace and the corresponding DDD collection trace, producing time-aligned feature sequences. The correlation coefficient is then calculated between the two sequences for each feature. We exclude instance types where one or both data sources exhibit no variation over the observation period, making the correlation undefined. The remaining 30 instance types form the analysis set.

Fig.~\ref{fig:feature-fidelity} presents the CDFs of the resulting correlation coefficients. All three features exhibit a distribution skewed toward positive values, indicating that DDD-derived features generally track the trends observed in actual instance data. Among the three, UR shows the strongest agreement with a median correlation of 0.90, with the majority of instance types yielding high positive correlations. SR and CUT exhibit weaker correlations with median values of 0.40 and 0.26 respectively, though their distributions still lean positive across most instance types. These results suggest that DDD-derived features, particularly UR, capture relevant availability patterns to contribute to predicting the state of actual running instances, though the fidelity varies across feature types.

\begin{figure}
    \centering
    \includegraphics[width=0.97\columnwidth]{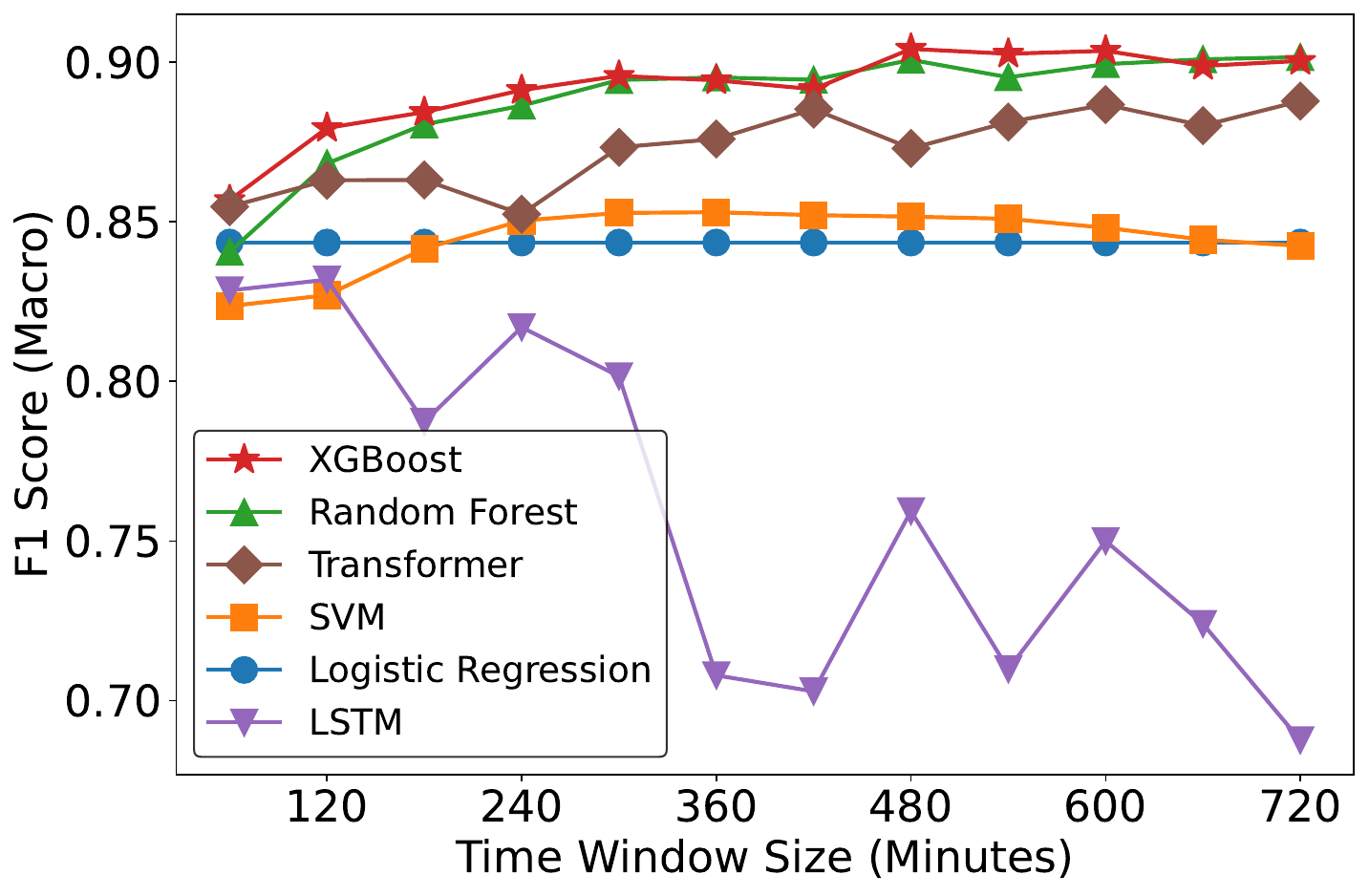}
    \caption{Availability modeling performance with DDD-derived features}
    \label{fig:eval-windowsize-model}
\end{figure}

\subsection{Availability Prediction with Extracted Features}\label{predict}

We evaluate the predictive capability of DDD-derived features across two tasks: classifying the current availability state of a spot node pool and forecasting whether the node pool will remain available over a prediction horizon of up to 60 minutes.

Fig.~\ref{fig:eval-windowsize-model} shows the effect of window size on current availability prediction, where the prediction horizon is 0 minutes. For each model, we select the feature combination yielding the highest F1-macro score at each window size and report the resulting performance. RF, XGBoost, and Transformer improve as the window size increases and stabilize beyond approximately 480 minutes, indicating diminishing returns from further extending the temporal context. LR and SVM show no significant change in their best performance as the window size increases. LSTM achieves its peak at a window size of 120 minutes and degrades progressively as the window grows. Rather than searching for an optimal window size, we select the window size that yields the best performance for each model individually and use these configurations to evaluate prediction over horizons ranging from 3 to 60 minutes.

\begin{figure*}[t]
    \centering
    \includegraphics[width=0.7\textwidth]{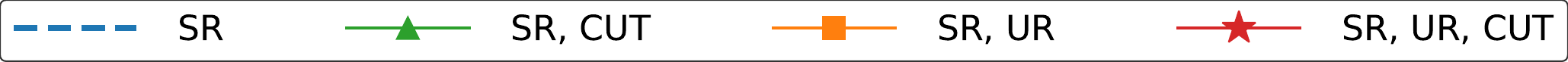}

    \subfloat[Logistic Regression]{
        \includegraphics[width=0.305\textwidth, trim=55 0 0 0]{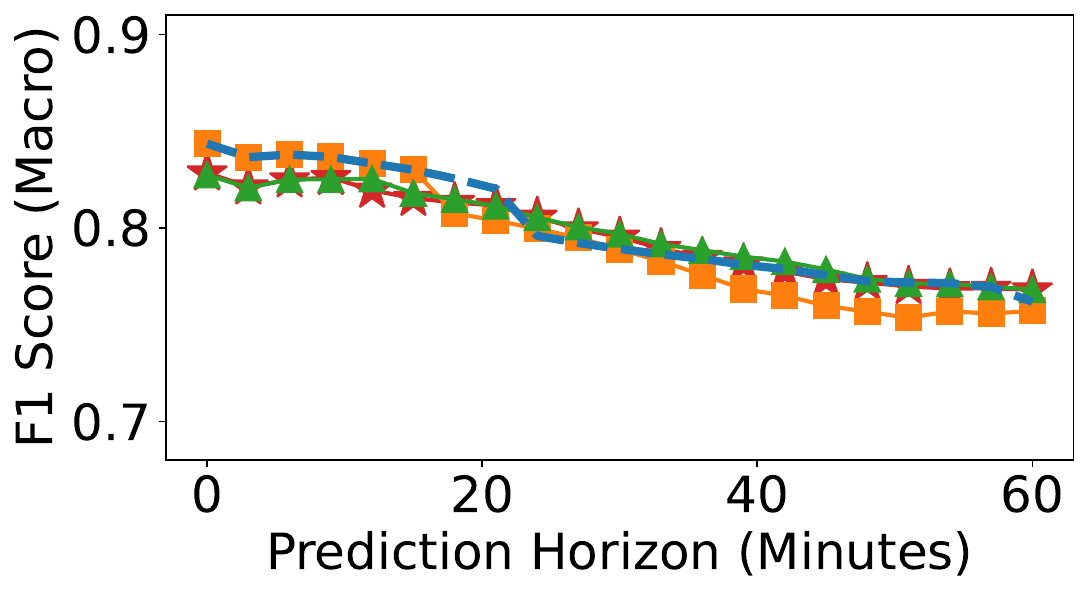}
        \label{fig:model-future-ddd-0}
    }%
    \subfloat[SVM]{
        \includegraphics[width=0.305\textwidth]{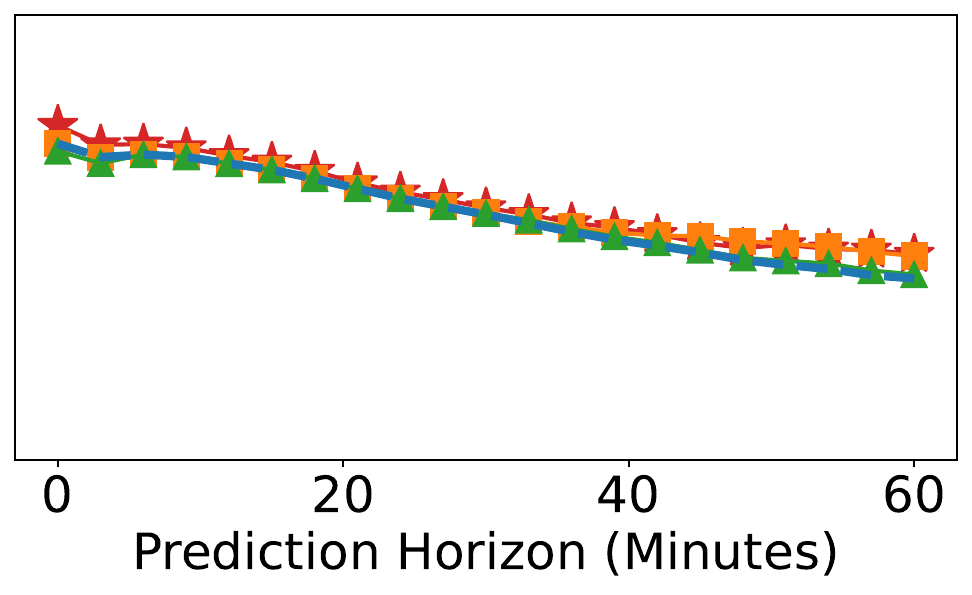}
        \label{fig:model-future-ddd-1}
    }%
    \subfloat[Random Forest]{
        \includegraphics[width=0.305\textwidth]{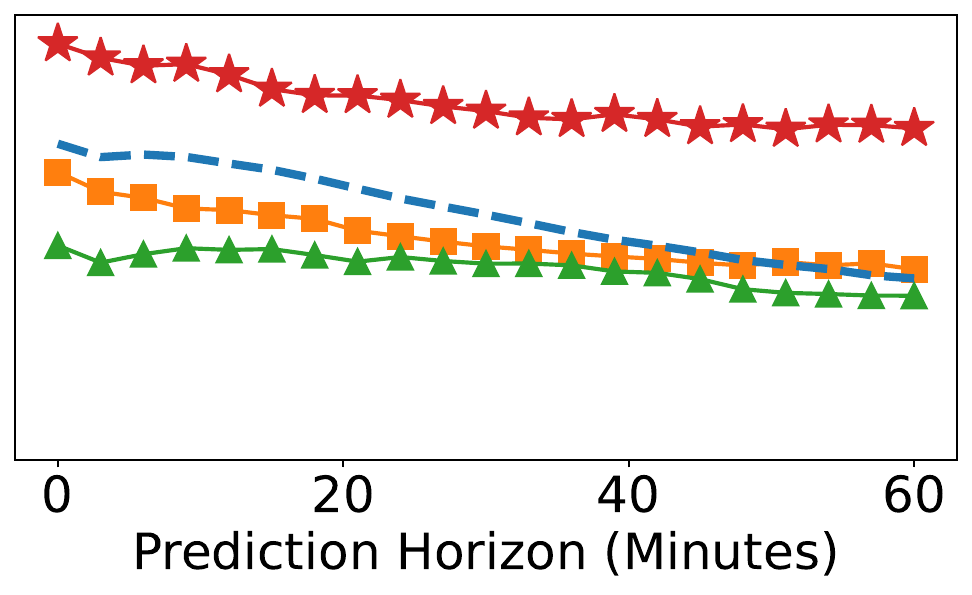}
        \label{fig:model-future-ddd-2}
    }%
    \\[-4pt]
    \subfloat[XGBoost]{
        \includegraphics[width=0.305\textwidth, trim=55 0 0 0]{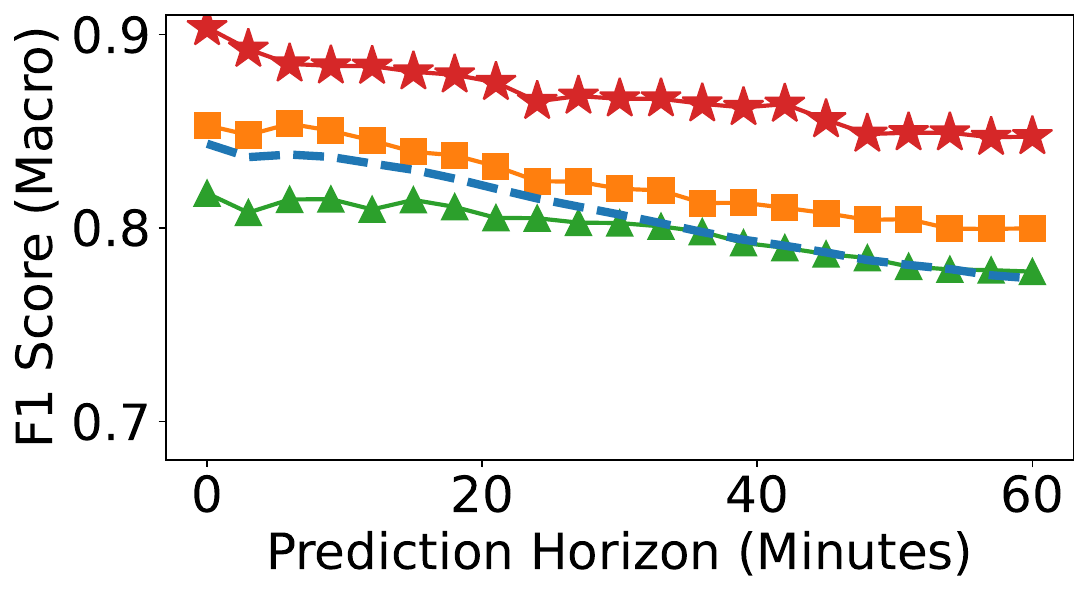}
        \label{fig:model-future-ddd-3}
    }%
    \subfloat[LSTM]{
        \includegraphics[width=0.305\textwidth]{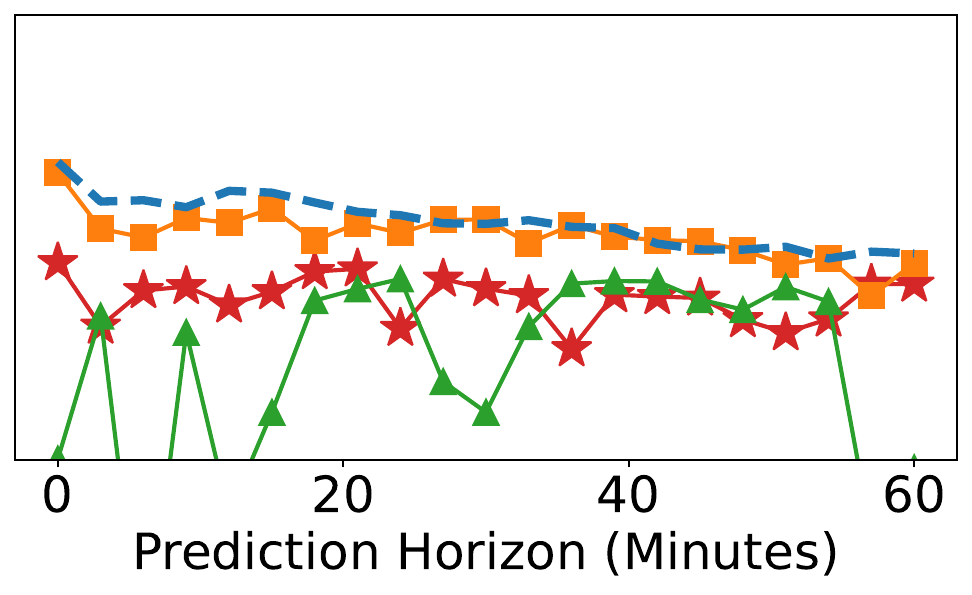}
        \label{fig:model-future-ddd-4}
    }%
    \subfloat[Transformer]{
        \includegraphics[width=0.305\textwidth]{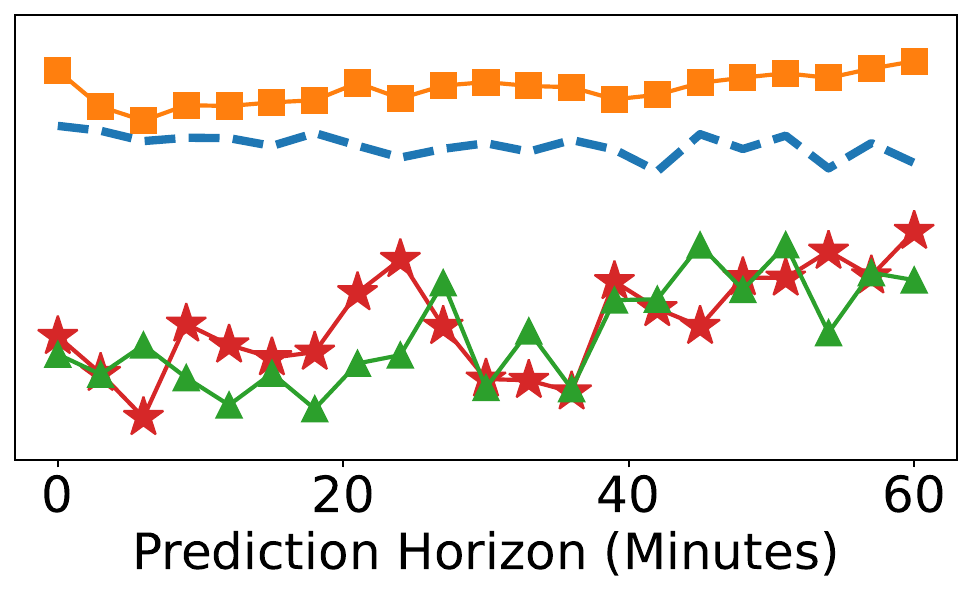}
        \label{fig:model-future-ddd-5}
    }%
    \caption{F1 macro score by prediction model and feature combination across prediction horizons}
    \label{fig:models-future}
\end{figure*}

Fig.~\ref{fig:models-future} presents the F1-macro scores of each model across prediction horizons from 3 to 60 minutes, evaluated with different feature combinations. Across all models, using SR alone yields consistent performance, confirming that the DDD success ratio itself serves as a strong baseline signal for availability prediction. Logistic Regression and Support Vector Machine show comparable or lower performance when additional features are included. For Random Forest and XGBoost, adding a single feature alongside SR does not improve performance, while combining both UR and CUT with SR achieves the highest scores, starting above 0.90 at a 3-minute horizon and maintaining approximately 0.85 at 60 minutes. This result reinforces the complementary relationship between UR and CUT discussed in Section~\ref{extracting-features}. LSTM shows limited performance gains despite its model complexity. With at most three input dimensions for a binary classification target, the task is too simple to benefit from recurrent sequence modeling, and UR and CUT, being temporally aggregated from SR, tend to act as noise when the model already learns temporal patterns from the input sequence. Transformer shows gradually improving performance as the prediction horizon increases with the SR+UR combination. A closer examination of the per-class F1 scores reveals that the F1 score for the unavailable class increases with longer horizons, indicating that the model learns to better capture upcoming unavailability events over extended prediction ranges. This suggests that Transformer may offer an advantage for longer-term availability prediction.

Despite these variations across models, the most consistent finding is that Random Forest and XGBoost paired with the full feature set achieve the strongest and most stable performance across all horizons. This indicates that well-designed features capturing the temporal characteristics of availability signals play a more critical role than model complexity for DDD-based availability prediction.

\subsection{Interruption Prediction}

To validate the practical utility of the DDD-based availability predictor, we conducted a trace-driven simulation that replays real spot availability data against a representative batch analytics workload. TPC-DS, a widely-used decision support benchmark~\cite{tpc-ds}, was chosen because its diverse mix of 99 analytical queries produces a realistic distribution of execution times for batch data processing on spot instances. We generated data at scale factor 300 and profiled all 99 queries on a Spark cluster of ten EC2 m5.large instances, yielding per-query execution times from 0.5 s to 661.5 s with a total workload duration of approximately 206 minutes.

The simulation replays the 24-hour DDD availability traces collected at three-minute intervals. To ensure a fair evaluation, the 68 instance types were split at the instance-type level into 75\% for training and 25\% for evaluation, preventing the trace of any evaluation instance type from being exposed to the model during training. To account for the effect of query ordering, each simulation was repeated five times with different random permutations of the query queue, and the reported metrics are averaged over these runs.

During each run the 99 profiled queries are executed sequentially over the full trace of an evaluation instance type. Queries proceed while all nodes in the pool remain available, and the progress of the running query is lost as soon as any node becomes unavailable. Three strategies are compared. Always Run serves as the baseline, launching the next queued query immediately without any prediction. Shortest Job First sorts the queue by ascending execution time so that shorter queries complete first, reducing the expected loss per interruption without relying on a prediction model. Predict-AR consults the DDD-trained XGBoost predictor at every collection cycle. When the model forecasts an upcoming unavailability, the strategy defers launching a new query for a duration equal to the prediction horizon, while leaving any already-running query undisturbed.

\begin{figure}
    \centering
    \includegraphics[width=0.97\columnwidth]{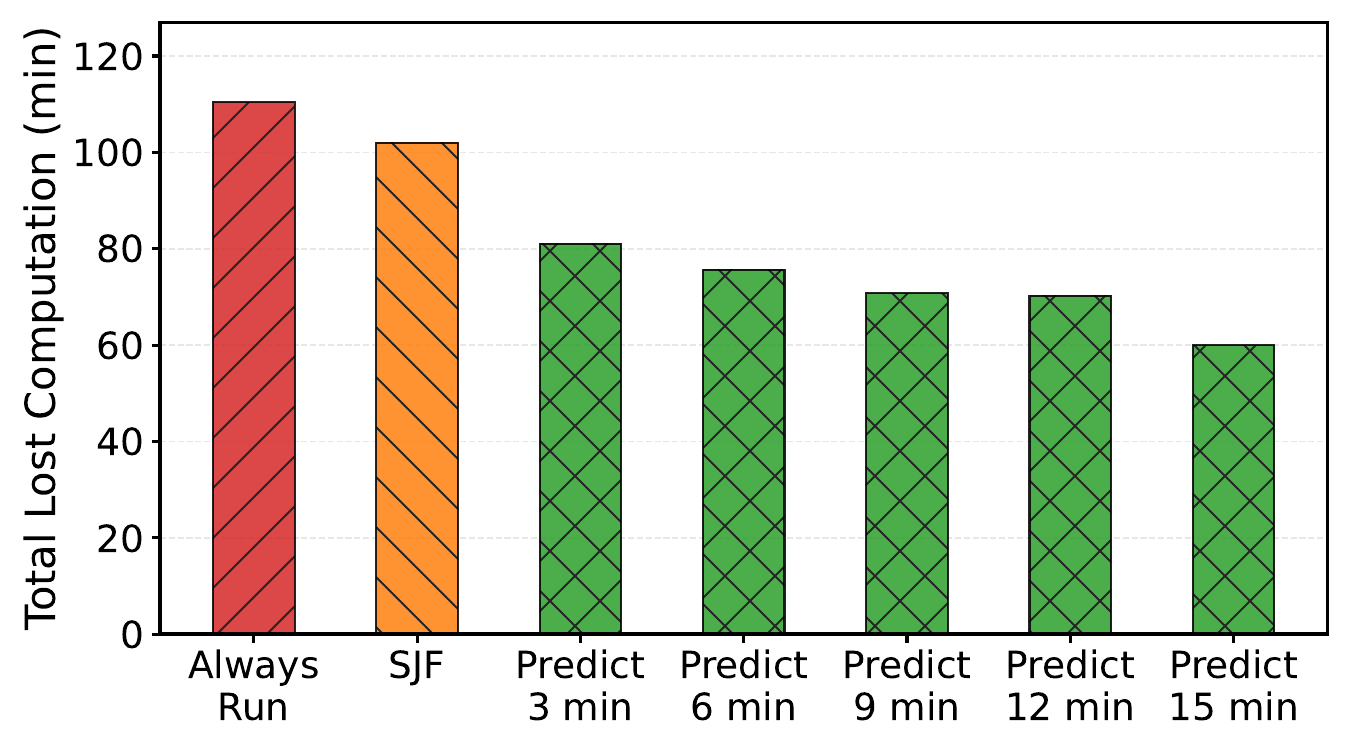}
    \caption{Total lost computation by strategy.}
    \label{fig:eval-simul}
\end{figure}

 Across the 17 evaluation instance types, Predict-AR reduced total lost computation by 27\% relative to Always Run when using the three-minute-ahead model and by up to 46\% with the fifteen-minute-ahead model, as shown in Fig.~\ref{fig:eval-simul}. However, false positives caused the strategy to defer job submission during periods that were in fact available, introducing idle time. As the prediction horizon lengthened, the reduction in lost computation grew while idle time increased. This result indicates a trade-off between loss avoidance and resource utilization.

\section{Discussion}

\textbf{Conservatism of DDD signals.}
As observed in the experimental results, DDD provides a conservative estimate of availability compared to the actual running instance count. This stems from the fact that capacity fluctuations affect the acceptance of new spot requests more readily than the maintenance of already running instances. From a practical standpoint, this conservatism is a favorable property. When a DDD request succeeds, the corresponding instance type is highly likely to be genuinely available, making the signal suitable as a lower-bound indicator for spot capacity.

\textbf{DDD as a provider-agnostic availability signal.}
Several cloud providers operate preemptible or spot-like pricing policies comparable to those of AWS and Azure. However, only a limited number of providers currently publish fine-grained stability or availability metrics for their spot offerings. Many providers expose no such indicators, leaving users without reliable means to assess instance-level availability before committing workloads. Because DDD relies solely on standard instance provisioning APIs rather than provider-specific availability endpoints, it can serve as a portable availability measurement method applicable to any provider that supports on-demand spot instance requests.

\textbf{Ethical considerations.}
Cloud providers impose per-hour or per-minute API rate limits to prevent service overload from excessive client requests. The DDD collection process involves repeated API calls at short intervals, and although all requests in this study were issued within the rate limits set by each provider, the potential impact of large-scale repeated provisioning requests on provider resource allocation systems warrants consideration. If collection is scaled to considerably larger deployments, conducting additional impact assessments would be prudent, including whether the collection activity itself influences the spot availability signals being measured.

\section{Conclusion and Future Work}
In this paper, we proposed DDD, a cost-efficient methodology for collecting cloud spot instance availability data through the provider's provisioning process without keeping instances running. We empirically validated that DDD signals closely reflect actual spot availability across diverse instance types and regions on both AWS and Azure, and defined availability features that capture complementary temporal aspects of spot dynamics. The proposed features enabled effective availability modeling and interruption prediction across multiple prediction horizons, and the integrated system architecture demonstrated that collection, feature computation, and prediction can operate within a unified pipeline.


For future work, we plan to incorporate an online learning pipeline with delayed labels, where actual interruption outcomes observed after prediction are fed back to update the model incrementally at each collection cycle, enabling the predictor to adapt to evolving availability patterns without full retraining. We also aim to integrate DDD signals with provider-supplied datasets, such as Spot Placement Scores and interruption frequency indicators, to enrich the feature space and further improve prediction accuracy. More broadly, we intend to explore methodologies that can more effectively leverage DDD-derived data for interruption prediction, pursuing both improved prediction performance and greater practical utility of DDD-based availability signals.

\section*{Acknowledgment}
This work was supported by Institute of Information \& communications Technology Planning \& Evaluation (IITP) grant funded by the Korea government (MSIT) (RS-2022-00144309 \& RS-2025-25441560 \& RS-2026-25492200)

\bibliographystyle{IEEEtran}
\bibliography{ddd}

\end{document}